\documentclass[twocolumn,prl,aps,showpacs]{revtex4}
\usepackage{graphicx}

\begin{document}

\title{\bf  Trapped Resonant Fermions above Superfluid Transition Temperature}

\author{Chi-Ho Cheng and Sung-Kit Yip}
\affiliation{Institute of Physics, Academia Sinica, Taipei,
Taiwan}

\date{\today}

\begin{abstract}
We investigate trapped resonant fermions with unequal populations
within the local density approximation above the superfluid
transition temperature. By tuning the attractive interaction
between fermions via Feshbach resonance, the system evolves from
weakly interacting fermi gas to strongly interacting fermi gas,
and finally becomes bose-fermi mixture. The density profiles of
fermions are examined and compared with experiments. We also point
out the simple relationships between the local density, the axial
density, and the gas pressure within the local density
approximation.
\end{abstract}


\maketitle


The experimental investigation of the crossover from the
Bardeen-Cooper-Schrieffer (BCS) state to the Bose-Einstein
Condensation (BEC) state induced by Feshbach resonance for trapped
fermionic atomic gases with equal \cite{greiner, zwierlein03,
bartenstein, bourdel, partridge05} and unequal populations
\cite{zwierlein1, partridge, zwierlein2} have attracted lots of
interests from the physics community. For equal populations, the
ground state of the system evolves from weak-coupling BCS to
strong-coupling BEC as the effective attraction between two
species of fermions becomes strong \cite{nozieres, demelo, ohashi,
carlson, perali1,stajic}. In the case of unequal populations of
the two fermion species, the system evolves from a normal state,
through a spatially inhomogeneous state(s) (like
Fulde-Ferrell-Larkin-Ovchinnikov (FFLO) state and/or phase
separation) in the weak-coupling regime to a bose-fermi mixture in
the strong-coupling regime
\cite{mizushima,pao1,sheehy,pao2,pieri,liu,silva1,silva2,parish}.

Although mean-field treatments can provide a qualitative picture,
inclusion of fluctuations \cite{nozieres} around the mean-field
solution is necessary to give results with more quantitative
agreement with experiments. For example, thermodynamics
\cite{kinast,chen} and density profiles \cite{perali1,stajic} at
finite temperature can be properly accounted for only when pair
fluctuations are included. Including these fluctuations
\cite{perali2,haussmann06} can also give a reasonable value for
the zero-temperature universal \cite{baker, heiselberg} parameter
$\beta$ at resonance, though it does not match exactly with the
Quantum Monte-Carlo results \cite{carlson, chang, astrakharchik}.

Till now, most of the studies are below the superfluid transition
temperature. However, above the transition temperature, the normal
fermion gas can still be strongly interacting and becomes a
mixture of bose and fermi gases in the strong coupling regime. The
system provides a good testing ground for strongly interacting
many-body theories.

On the other hand, the thermometry of strongly interacting fermi
gas is also important and difficult in experiments \cite{kinast}.
In the case of non-interacting fermions, the temperature can be
simply measured by fitting the density profile of trapped fermions
with the Thomas-Fermi distribution. Thermometry of strongly
interacting fermions with equal populations requires non-trivial
fitting procedure on the theoretically generated density profiles
\cite{chen}. In the case of unequal populations of two species of
fermions, the wing of the majority component (excess fermion)
becomes non-interacting due to the absence of the minority
component, and thus can be served as a good thermometer
\cite{zwierlein2}.

In this paper, the Nozieres-Schmitt-Rink (NSR) formalism
\cite{nozieres} is adopted since it is the simplest theory that
can continuously bring both the BCS and BEC regimes respectively
for weak and strong coupling limits together at finite
temperatures, and also gives a good estimate of the universal
parameter in the unitarity limit as stated above. The harmonic
trapping potential is considered within the local density
approximation (LDA).

We found that the Hartree-Fock approximation is recovered in the
weak-coupling regime, but it is qualitatively different from the
NSR treatment if the system evolves into a strongly interacting
fermi gas. In the strong coupling limit, the system becomes a
bose-fermi mixture. An effective repulsion between the bosonic
molecules and fermions emerges from the NSR theory. Not so
surprisingly, a similar repulsive effect also exists in the
intermediate coupling regime from the NSR theory although no bound
pair is formed. The density profiles are obtained, comparison and
implication to the recent experiment are then discussed.


We adopted the implementation of the NSR formalism by the
functional integral method \cite{demelo}. The action of our system
is written as
\begin{eqnarray} \label{action}
S &=& \int d\tau d\vec{r} \{ \sum_\sigma \overline\psi_\sigma
(\partial_\tau -\nabla^2/2m - \mu_\sigma + V(\vec r)) \psi_\sigma
\nonumber
\\ && + g \overline\psi_\uparrow \overline\psi_\downarrow
\psi_\downarrow \psi_\uparrow  \}
\end{eqnarray}
with attractive coupling constant $g<0$. $\sigma$ runs over up and
down spin species, and $\psi_\sigma\equiv\psi_\sigma(\vec r,\tau)$
are the fermionic fields. Two chemical potentials $\mu_\sigma$ fix
the fermion number densities inside the trapping potential $V(\vec
r)$.

By introducing the Hubbard-Stratonovich field $\Delta(\vec
r,\tau)$ coupled to the pairing field $\overline\psi_\uparrow
\overline\psi_\downarrow$, and then integrating out the fermion
fields $\psi_\sigma$, we get the effective action in terms of
$\Delta(\vec r,\tau)$. Above the superfluid transition temperature
with vanishing saddle point $\Delta_0 = 0$, we expand the
effective action around the saddle point up to the Gaussian level
in $\Delta(\vec q, i\!p_n)$, which gives
\begin{eqnarray}
S_{\rm eff} = S_0 + \sum_{\vec q, i\!p_n} \Gamma^{-1}(\vec q,
i\!p_n) |\Delta(\vec q, i\!p_n)|^2
\end{eqnarray}
where
\begin{eqnarray} \label{gamma}
\Gamma^{-1}(\vec q, i\!p_n) &=& \frac{1}{L^3}\sum_{\vec k} \left(
\frac{1-n_{\rm F}(\xi_{\vec q/2 + \vec k, \uparrow}) - n_{\rm F
}(\xi_{\vec q/2 - \vec k, \downarrow})}{i\!p_n - \xi_{\vec q/2 +
\vec k, \uparrow} - \xi_{\vec q/2 - \vec k, \downarrow}} \right.
\nonumber
\\ && \left. + \frac{1}{2 \epsilon_{\vec k}} \right)  - \frac{m}{4\pi a}
\end{eqnarray}
with $\epsilon_{\vec k}=k^2/2m$, $\xi_{\vec k, \sigma} \equiv
\epsilon_{\vec k}-\mu_\sigma + V(\vec r)$, and $n_{\rm F}$ is the
Fermi distribution function. The $s$-wave scattering length $a$
defined by two-body low-energy scattering via $m/4\pi a = 1/g +
L^{-3}\sum_{\vec k}1/2\epsilon_{\vec k}$ is to regulate the
ultraviolet divergence. Following Nozi\`{e}res and Schmitt-Rink
\cite{nozieres}, we can rewrite the free energy
\begin{eqnarray}
F = F_0 - \frac{1}{L^3}\sum_{\vec q}\int_{-\infty}^{+\infty}
\frac{d\omega}{\pi} n_{\rm B}(\omega) \delta(\vec q, \omega)
\end{eqnarray}
in terms of the phase shift $\delta(\vec q, \omega)$ defined by
$\Gamma(\vec q, \omega+i0^+)=|\Gamma(\vec q,
\omega)|\exp(i\delta(\vec q, \omega))$. $F_0$ is the free energy
of free fermions and $n_{\rm B}$ is the Bose distribution
function. The number equations are given by
\begin{eqnarray} \label{number}
n_\sigma(\vec r) = n^0_\sigma(\vec r) + \frac{1}{L^3}\sum_{\vec
q}\int_{-\infty}^{+\infty}\frac{d\omega}{\pi} n_{\rm B}(\omega)
\frac{\partial}{\partial\mu_\sigma}\delta(\vec q, \omega)
\end{eqnarray}
with the bare occupation $n^0_\sigma(\vec r)=L^{-3}\sum_{\vec k}
n_{\rm F}(\xi_{\vec k, \sigma})$.

Our density profiles are calculated from Eq.(\ref{number}). To
clarify the physics, we first examine two limiting cases. In the
extreme weak coupling limit with $a<0$ and $k_{\rm F}|a| \ll 1$,
$\Re{\rm e} \Gamma^{-1} \simeq -m/4\pi a $, with the first term in
Eq.(\ref{gamma}) contributing a small imaginary part. Hence
\begin{eqnarray}
\delta(\vec q,\omega) &\simeq& \frac{4\pi a}{m} \Im{\rm m}
\Gamma^{-1}(\vec q,\omega)
\end{eqnarray}
and the second term in Eq.(\ref{number}) becomes
\begin{eqnarray}
&&\frac{4\pi a}{m} \frac{\partial}{\partial \mu_\sigma}
\frac{1}{L^3}\sum_{\vec
q}\int_{-\infty}^{+\infty}\frac{d\omega}{\pi} n_{\rm
B}(\omega)\Im{\rm m} \Gamma^{-1}(\vec q,\omega)
 \nonumber \\ &=& -\frac{4\pi a}{m}
\frac{\partial}{\partial \mu_\sigma} \frac{1}{L^6}\sum_{\vec
q,\vec k}
n_{\rm B}(\xi_{\vec q/2 + \vec k, \uparrow} + \xi_{\vec q/2 - \vec k, \downarrow}) \nonumber \\
&&\times(1-n_{\rm F}(\xi_{\vec q/2 + \vec k, \uparrow})-n_{\rm
F}(\xi_{\vec q/2 - \vec k, \downarrow}))
 \nonumber \\ &=&
-\frac{4\pi a}{m} \frac{\partial}{\partial \mu_\sigma}
\frac{1}{L^6}\sum_{\vec q,\vec k} n_{\rm F}(\xi_{\vec q/2 + \vec
k, \uparrow})n_{\rm F}(\xi_{\vec q/2 - \vec k, \downarrow})
 \nonumber \\ &=&
-\frac{4\pi a}{m} n^0_{-\sigma}(\vec r)\frac{\partial}{\partial
\mu_\sigma}n^0_{\sigma}(\vec r)
\end{eqnarray}
where the bose distribution function is eliminated by the
identity, $n_{\rm B}(x+y)(1-n_{\rm F}(x)-n_{\rm F}(y))=n_{\rm
F}(x)n_{\rm F}(y)$. The occupation in Eq.(\ref{number}) reduces to
\begin{eqnarray} \label{number-hf}
n_\sigma(\vec r) &\simeq& n^0_\sigma(\vec r) -\frac{4\pi a}{m}
n^0_{-\sigma}(\vec r)\frac{\partial}{\partial
\mu_\sigma}n^0_{\sigma}(\vec r) \nonumber \\
&\simeq& \frac{1}{L^3}\sum_{\vec k} n_{\rm F}(\xi_{\vec k,
\sigma}+\frac{4\pi a}{m} n^0_{-\sigma}(\vec r))
\end{eqnarray}
The expression is equivalent to the first-order Hartree-Fock (HF)
approximation to the original Hamiltonian in Eq.(\ref{action})
with the effective coupling constant $4\pi a/m$ instead of the
bare $g$.

\begin{figure*}[tbh]
\begin{center}
\includegraphics[width=5in]{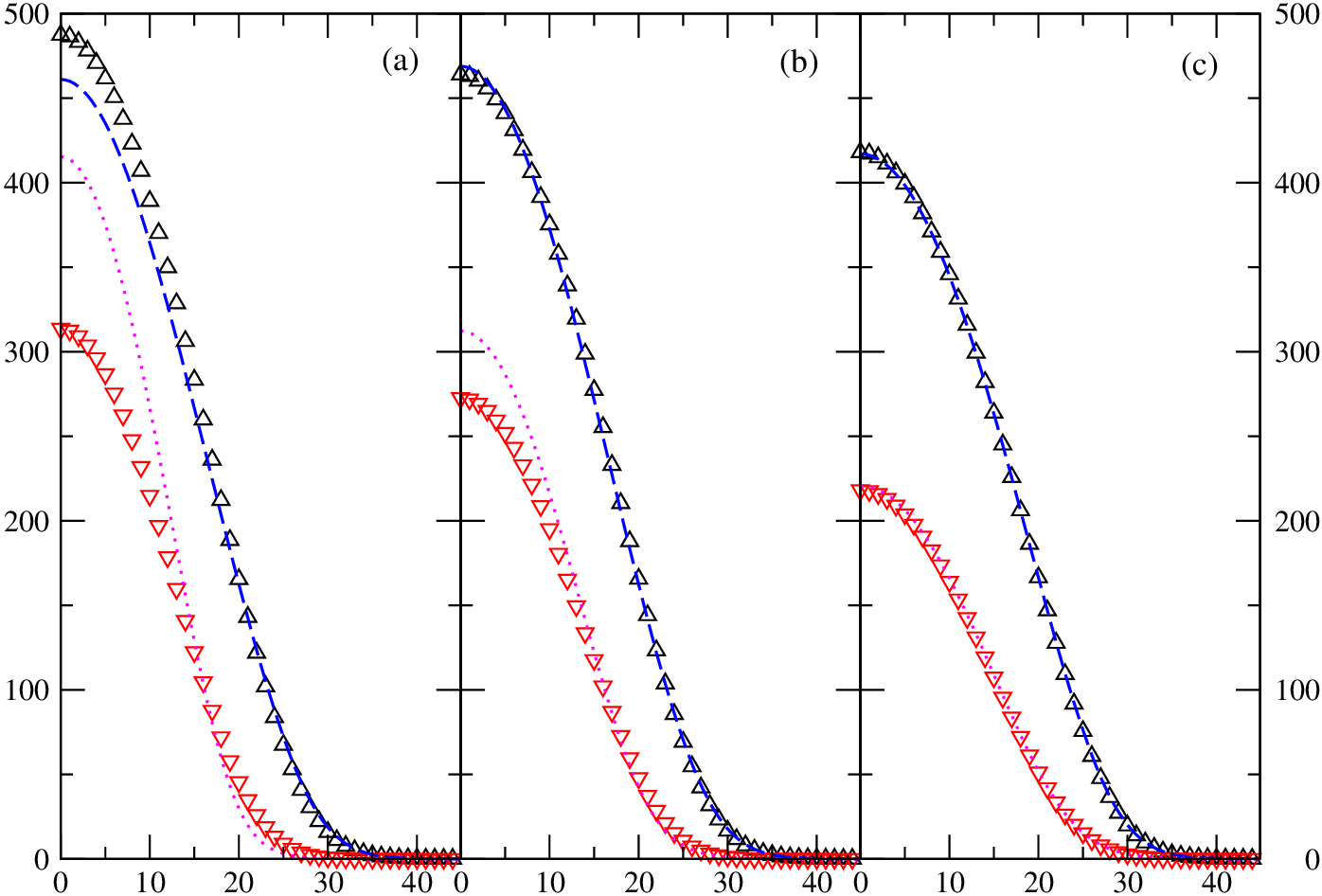}
\end{center}
 \caption{(Color online) Comparison between density profiles of NSR
 fluctuation theory and that of Hartree-Fock approximation at weak coupling regime.
 Labels of horizontal and vertical axes are respectively
 dimensionless $r/\ell$ and $n(r)\ell^3$. Up-spin and down-spin profiles calculated from
 fluctuation theory are represented by up-triangle and down-triangle, respectively.
 The dashed and dotted lines show the Hartree-Fock approximation to
 the up-spin and down-spin fermion profiles, respectively. The
 parameter range is chosen as $N_\uparrow = 1.5 \times 10^7$,
 $N_\downarrow = 5 \times 10^6$, and $T/T_{\rm F} = 0.2$.
 The interaction parameters $-1/k_{\rm F}a$ are respectively (a) 0.5 (b) 1.0 (c) 10.0 }
 \label{compare.eps}
\vspace{5pt}
\end{figure*}

In Fig. \ref{compare.eps}, we plot both the density profiles of
the NSR and HF theories. At extremely small $k_{\rm F} |a|$, as
shown in Fig. \ref{compare.eps}(c), two theories coincide with
each other. As the interaction parameter increases, $\Re{\rm
e}\Gamma^{-1} > -m/4\pi a$ due to the contribution from the first
term in Eq.(\ref{gamma}). The HF approximation overestimates the
effective attractive interaction.  As shown in Fig.
\ref{compare.eps}(b), the HF results for the minority component
lies above our fluctuation theory results near the trap center.
The deviations between these two theories are larger for the
minority than the majority component, since the minority component
actually feels a larger change in effective particle energy (see
the argument of the function $n_{\rm F}$ in Eq.(\ref{number-hf})).

Upon further increase in $k_{\rm F} |a|$ (Fig.
\ref{compare.eps}(a)), the NSR theory shows even qualitative
disagreement with the HF results. It can be seen that the majority
profile from NSR theory lies {\em above} the HF results near the
trap center. The reason of the disagreement will be discussed
further below.

In the strong coupling limit where $a>0$, the two-particle vertex
function $\Gamma(\vec q, \omega)$ acquires a discrete pole
$\omega=\omega_{\rm b}(\vec q)$ representing the bound state of
fermion-pair. Around the pole, we can write
\begin{eqnarray} \label{pole}
\Gamma(\vec q, \omega) = \frac{R}{-\omega + \omega_{\rm b}(\vec
q)}
\end{eqnarray}
with small $\vec q$, and the dispersion $\omega_{\rm b}(\vec
q)=c_0 + c_1 q^2$. Eq.(\ref{pole}) can also be written as
\begin{eqnarray} \label{pole2}
\omega_{\rm b}(\vec q) - \omega =  R \Gamma^{-1}(\vec q, \omega)
\end{eqnarray}
The coefficient $c_0$ can be determined from putting $\vec q = 0$,
$\omega = c_0$ into Eq.(\ref{pole2}), which gives $\Gamma^{-1}(0,
c_0)=0$, {\it i.e.},
\begin{eqnarray}
\frac{1}{L^3}\sum_{\vec k} \left( \frac{1-n_{\rm F}(\xi_{\vec k,
\uparrow}) - n_{\rm F }(\xi_{\vec k, \downarrow})}{c_0 - \xi_{\vec
k, \uparrow} - \xi_{\vec k, \downarrow}} + \frac{1}{2
\epsilon_{\vec k}} \right) = \frac{m}{4\pi a}
\end{eqnarray}
We rewrite this formula as
\begin{widetext}
\begin{eqnarray} \label{c0}
\frac{1}{L^3}\sum_{\vec k} \left( \frac{1}{(c_0+\mu_\uparrow +
\mu_\downarrow -2V(\vec r)) - 2\epsilon_{\vec k}} + \frac{1}{2
\epsilon_{\vec k}} \right) - \frac{1}{L^3}\sum_{\vec k}
\frac{n_{\rm F}(\xi_{\vec k, \uparrow})+n_{\rm F}(\xi_{\vec k,
\downarrow})}{(c_0+\mu_\uparrow + \mu_\downarrow -2V(\vec r)) -
2\epsilon_{\vec k}}   = \frac{m}{4\pi a}
\end{eqnarray}
\end{widetext}
With equal population, the chemical potential of the bound state
$\mu_\uparrow=\mu_\downarrow < 0$, the second term of the L.H.S.
of Eq.(\ref{c0}) vanishes at temperature $k_{\rm B}T \ll
|\mu_{\uparrow,\downarrow}|$. There we can obtain $c_0 =
\epsilon_{\rm b} - \mu_\uparrow - \mu_\downarrow + 2V(\vec r)$,
with the binding energy $\epsilon_{\rm b} = -1/ma^2 < 0$. In our
case of unequal populations, the chemical potential of excess
(unpaired) fermions $\mu_\uparrow > 0$ and that of the bound state
$\mu_\downarrow < 0$. For temperatures $k_{\rm B}T \ll
|\epsilon_{\rm b}|$, we can substitute the above $c_0$ of equal
population into the second term of the L.H.S. of Eq.(\ref{c0}),
ignore $\epsilon_{\vec k}$ in the denominator, and obtain for this
term $-ma^2 \delta n^0_{\rm f}(\vec r)$, where $\delta n^0_{\rm
f}(\vec r)\equiv L^{-3}\sum_{\vec k}n_{\rm F}(\xi_{\vec k,
\uparrow})$ can be interpreted as the occupation due to bare
excess fermion. Eq.(\ref{c0}) then gives
\begin{eqnarray}
&& \frac{1}{L^3}\sum_{\vec k} \left( \frac{1}{(c_0+\mu_\uparrow +
\mu_\downarrow -2V(\vec r)) - 2\epsilon_{\vec k}} + \frac{1}{2
\epsilon_{\vec k}} \right) \nonumber \\
&=& \frac{m}{4\pi a} - ma^2 \delta n^0_{\rm f}(\vec r)
\end{eqnarray}
hence we get
\begin{eqnarray}
c_0 &=& -\frac{1}{ma^2}(1- 4\pi a^3 \delta n^0_{\rm f}(\vec r))^2
-\mu_\uparrow - \mu_\downarrow + 2V(\vec r) \nonumber \\
&\simeq& \epsilon_{\rm b} + g_{\rm bf}\delta n^0_{\rm f}(\vec r) -
\mu_\uparrow - \mu_\downarrow + 2V(\vec r)
\end{eqnarray}
with $g_{\rm bf}\equiv 8\pi a/m$.

Similarly, $R$ can be determined by putting $\vec q=0$ and taking
the limit $\omega \rightarrow c_0$ in Eq.(\ref{pole2}),
\begin{eqnarray} \label{r0}
R^{-1} &=& \lim_{\omega \rightarrow c_0}
\frac{\Gamma^{-1}(0,\omega)}{c_0 - \omega}  \nonumber \\
&=& \frac{1}{L^3}\sum_{\vec k} \frac{1-n_{\rm F}(\xi_{\vec
k,\uparrow})-n_{\rm F}(\xi_{\vec k,\downarrow})}{(\epsilon_{\rm
b}+g_{\rm bf}\delta n^0_{\rm f}(\vec r) -2\epsilon_{\vec k} )^2}
\nonumber \\
&\simeq& \frac{m^2 a}{8\pi}\left( 1 - 4\pi a^3 \delta n^0_{\rm
f}(\vec r) \right)
\end{eqnarray}

By noticing that, in the denominator of the first term in
Eq.(\ref{gamma}),  the $q^2$-term can be combined with $i\!p_n$,
we found $\Gamma^{-1}(\vec q, \omega) = \Gamma^{-1}(0, \omega -
q^2/4m) + O(q^4)$. Therefore $c_1 = 1/4m$.

In the strong coupling regime, the low-energy effective action can
then be written as
\begin{eqnarray}
S_{\rm eff} = S_0 + \sum_{\vec q, i\!p_n} (-i\!p_n + \omega_{\rm
b}(\vec q)) \overline\phi(\vec q, i\!p_n)  \phi(\vec q, i\!p_n)
\end{eqnarray}
where we identify the boson field $\phi(\vec q, i\!p_n)\equiv
R^{-1/2}\Delta(\vec q, i\!p_n)$, and $\omega_{\rm b}(\vec q)\equiv
\epsilon_{\rm b} + g_{\rm bf}\delta n^0_{\rm f}(\vec r) -
\mu_\uparrow - \mu_\downarrow + 2V(\vec r) + q^2/4m$. The phase
shift across the pole $\omega = \omega_{\rm b}(\vec q)$ jumps from
$0$ to $\pi$, {\it i.e.}, $\delta(\vec q, \omega) = \pi \theta(
\omega_{\rm b}(\vec q) - \omega)$. Noticing that
\begin{eqnarray}
\frac{\partial}{\partial \mu_\uparrow}\delta(\vec q, \omega) &=&
\pi \delta(\omega - \omega_{\rm b}(\vec q) ) \left(1 - g_{\rm bf}
\frac{\partial}{\partial
\mu_\uparrow}\delta n_{\rm f}^0(\vec r)\right)  \\
\frac{\partial}{\partial \mu_\downarrow}\delta(\vec q, \omega) &=&
\pi \delta(\omega - \omega_{\rm b}(\vec q)) \  ,
\end{eqnarray}
and then the number equation in Eq.(\ref{number}) gives
\begin{eqnarray} \label{nup}
n_\uparrow(\vec r) &=& \delta n^0_{\rm f}(\vec r) + \left(n_{\rm
b}(\vec r) - g_{\rm bf}n_{\rm
b}\frac{\partial}{\partial\mu_\uparrow}\delta n^0_{\rm f}(\vec r)
\right) \nonumber \\
&\simeq& \delta n_{\rm f}(\vec r) + n_{\rm b}(\vec r)
\end{eqnarray}
and
\begin{eqnarray} \label{ndown}
n_\downarrow(\vec r) = n_{\rm b}(\vec r)
\end{eqnarray}
with $\delta n_{\rm f}(\vec r)=L^{-3}\sum_{\vec k}n_{\rm
F}(\xi_{\vec k, \uparrow}+g_{\rm bf}n_{\rm b}(\vec r))$ and
$n_{\rm b}(\vec r)=L^{-3}\sum_{\vec q}n_{\rm B}(\omega_{\rm
b}(\vec q ))$. Eqs.(\ref{nup})-(\ref{ndown}) are the number
equations in strong coupling limit. The effective interaction
characterized by $g_{\rm bf}$ between the bound pairs and residue
fermions are already captured in the Gaussian level. The same
results were also obtained by a low density zero-temperature
expansion by one of the authors \cite{yip} and self-consistent
Bogoliubov-de-Gennes equations by Pieri and Strinati \cite{pieri},
but this interaction was missing in the treatment of Liu and Hu
\cite{liu}. Note also that, in NSR formalism, $n_{\rm b}(\vec r)$
and $\delta n_{\rm f}(\vec r)$ are not determined
self-consistently. Instead, $n_{\rm b}(\vec r)$ relies (via
$\omega_{\rm b}(\vec q)$) only on the bare $\delta n^0_{\rm
f}(\vec r)$, and $\delta n_{\rm f}(\vec r)$ depends then on
$n_{\rm b}(\vec r)$.


With the LDA, the density profile of an anisotropic harmonic trap
can be mapped to a corresponding isotropic one by rescaling the
radial distances for the different axes. In the following, we
consider only the isotropic case.

\begin{figure*}[tbh]
\begin{center}
\includegraphics[width=6.5in]{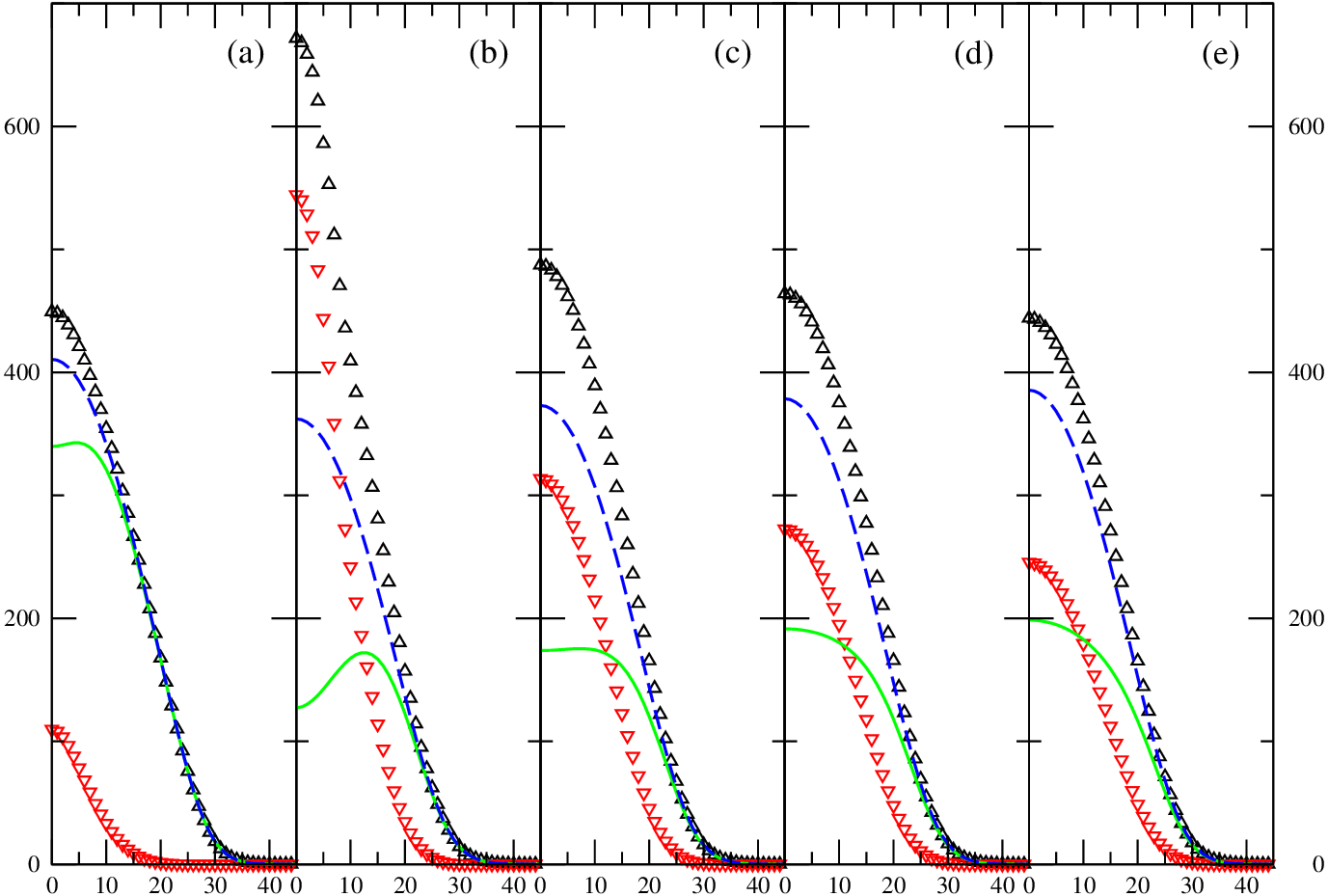}
\end{center}
 \caption{(Color online) Density profiles of fermions of spin up and
 down. Labels of horizontal and vertical axes are respectively
 dimensionless $r/\ell$ and $n(r)\ell^3$. Up-spin, down-spin, and excess fermion are represented by
 up-triangle, down-triangle, and solid line, respectively. The dashed lines shows the
 trapped non-interacting fermion fit to the wing of up-spin fermion
 profile.
 The dimensionless trap length $k_{\rm F}\ell=28$, and the temperature $T/T_{\rm F} = 0.2$.
 The total populations of up-spin fermions $N_\uparrow = 1.5 \times 10^7$.
 In strong coupling regime (a), the interaction parameters $-1/k_{\rm F}a = -2.0$,
 the total population of down-spin fermions $N_\downarrow = 5 \times 10^5$ with the population imbalance $\delta=93\% $.
 In weak coupling regime, $-1/k_{\rm F}a$ are respectively
 (b) 0.1 (c) 0.5 (d) 1.0 (e) 2.0 , $N_\downarrow = 5 \times 10^6$ with $\delta=50\% $. }
 \label{density.eps}
\end{figure*}

The density profiles of fermions of up and down-spin for different
interaction parameter $-1/k_{\rm F}a$ are shown in Fig.
\ref{density.eps}. The total populations of the majority species
is $1.5 \times 10^7$ with the population imbalances $50\%$, $T/T_F
= 0.2$ in  the weak-coupling regimes of Fig.
\ref{density.eps}(b)-(e).  These parameters are chosen to be close
to those in an experiment of MIT \cite{zwierlein2}.
 As the
interaction parameter becomes stronger, the two species of
fermions are strongly interacting with each other and the local
density at the trap center rises. The excess fermions defined by
the difference between the local densities of up and down-spin
fermions are repelled from the trap center. Thomas-Fermi fit to
the wing of the majority profile are insensitive to the
interaction parameter although the fermions density profile can be
strongly deformed. It is an evidence to the reliability of the
thermometry from the excess fermions. Because of the attractive
interaction from down-spin fermion, the profile of the
Thomas-Fermi fit always lies below that of up-spin fermion. When
we increase $k_{\rm F}a$ further close to the unitarity limit
(Fig. \ref{density.eps}(b)), there is a dramatic change in density
profiles. The bound pairs are close to being bose-condensed. It
explains why both species strongly peak near the trap center and
the density profiles behave qualitatively differently from that of
the HF results, as discussed above. The case of strong coupling
regime is shown in Fig. \ref{density.eps}(a) where a large
population imbalance was chosen to avoid the bose-condensation of
the bound pairs.

Near the resonance, the NSR theory may break down due to the
unphysical (non-positive definite) susceptibility matrix $\partial
n_\sigma/\partial \mu_{\sigma'}$ \cite{parish}. We checked that
the susceptibility matrix is positive definite with the same
parameters used in Fig. \ref{density.eps}. The NSR theory does not
break down at this point in our problem.

\begin{figure*}[tbh]
\begin{center}
\includegraphics[width=6.5in]{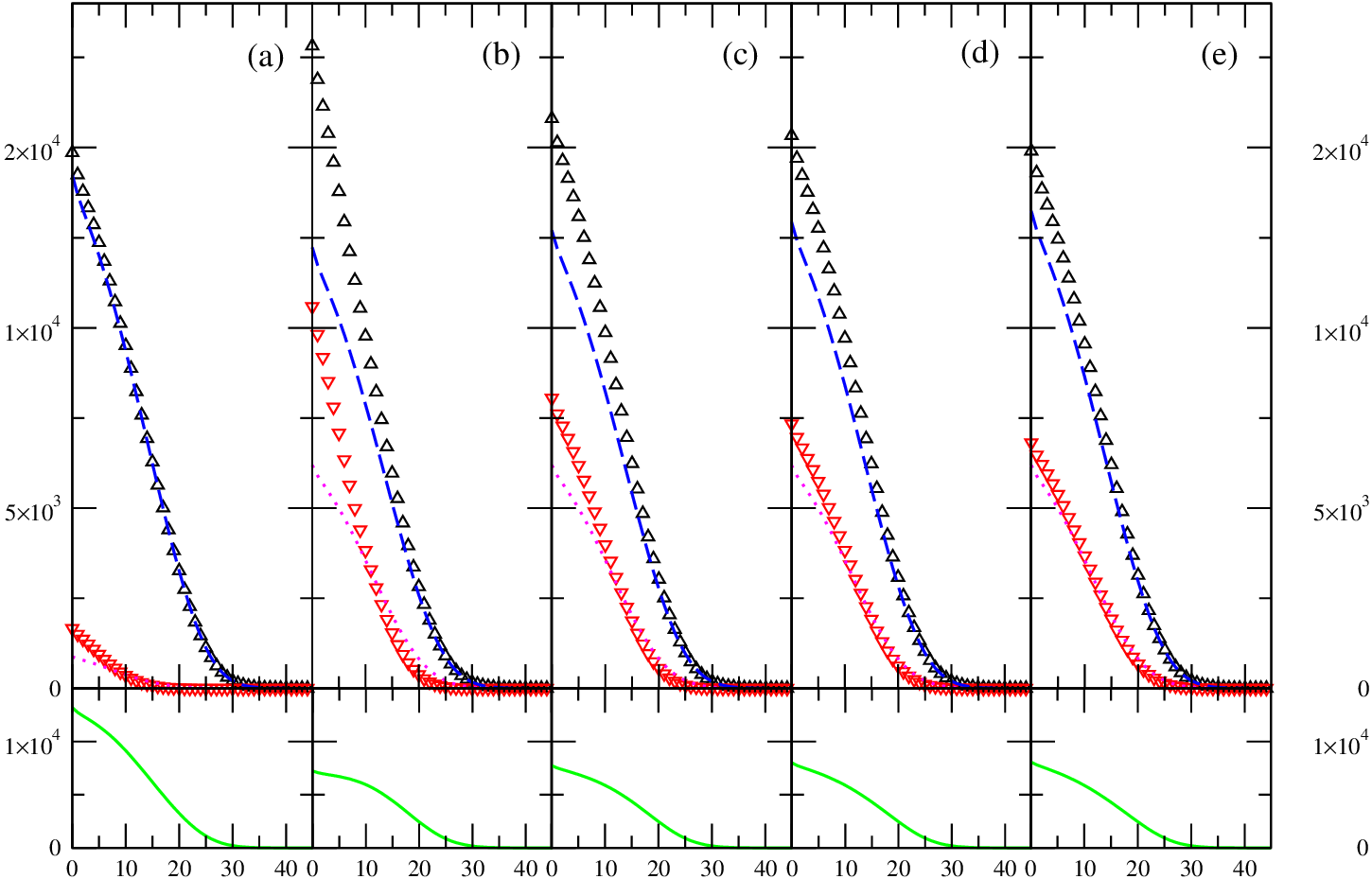}
\end{center}
 \caption{(Color online) Column density profiles of fermions of spin up and
 down. Labels of horizontal and vertical axes are respectively
 dimensionless $\rho/\ell$ and $n_{\rm col}(\rho)\ell^2$. Up-spin, down-spin, and excess fermion are represented by
 up-triangle, down-triangle, and solid line (lower panels)
 respectively. The dashed line shows the
 trapped non-interacting fermion fit to the wing of spin-up
 fermion profile. The dotted line is the non-interacting
 down-spin fermion profile at the same temperature, total
fermion numbers and trap length. Parameters are the same as in Fig
\ref{density.eps}. }
 \label{column.eps}
\end{figure*}

To compare our result more directly with experiments (see Fig. 1
in Ref. \cite{zwierlein2}), the column densities are shown in Fig.
\ref{column.eps}. The column density $n_{\rm col}(\rho)$ can be
obtained from the local density $n(r)$ by

\begin{eqnarray} \label{transform-column}
n_{\rm col}(\rho) \equiv  \int_{-\infty}^\infty dx  n(r) =
\int_\rho^\infty dr \frac{2r}{\sqrt{r^2-\rho^2}} n(r)
\end{eqnarray}

We also plot the Thomas-Fermi fit to the wing of the up-spin
fermions, and the non-interacting fermions profile with the the
same total population and temperature as that of the down-spin
fermions . Because of strong attraction around the trap center,
the interacting profiles always lie above the non-interacting one
near the center.


Compared with a recent experiment \cite{zwierlein2}, we found only
some qualitative similarities  but also rather significant
disagreements. The cloud sizes are found to be smaller than their
non-interacting counterparts, as in the experiments. However,
experimentally the density profiles deviate from the
non-interacting ones mainly for the majority cloud but not so much
for the minority cloud, whereas we found significant effects of
the attractive interaction on both. Experimentally, the density
profile of the majority particles actually lies {\em below} the
finite temperature non-interacting fit of the wings at where the
minority component appears, though this deficit decreases when $1/
k_{\rm F} a$ increases. Our calculated results give a density
profile {\em above} this fit, as expected from attractive
interaction between the particles. With increasing $1/ k_{\rm F}
a$, there is a larger and larger enhancement of the actual density
profile over the fit to the wings. A possible source of
disagreement is that our current understanding of the expansion
dynamics is not sufficiently quantitative, so that the scaling
factors for the interacting gas near the center versus that of the
non-interacting particles near the wing are different from what
was used in Ref. \cite{zwierlein2}.

For the minority component, our density profile shows strong
effects of the effective attractive interaction, so that it peaks
much more strongly near the trap center than a corresponding
non-interacting distribution as seen in Fig. \ref{column.eps}.
(whereas experimentally it can be fitted well by the
non-interacting profile).  We found that, even above the
transition temperature as it is here, the difference between the
majority and minority components ({\it i.e.}, the excess fermion
density) seems to decrease when $1 / k_{\rm F} a$ increases.  This
trend already exists in the HF results (see Fig. \ref{compare.eps}
and discussions above). It can also be understood as an effective
repulsive interaction between the bosons on the excess fermions,
so that the latter are pushed out from the trap center ({\it c.f.}
Fig. \ref{density.eps}, especially \ref{density.eps}(a) and (b)),
though this effect is not so evident in the column density profile
due to the integration.

In the following, we want to point out that, for a harmonic trap
within the local density approximation, the axial density (from
integration of the local density) directly provides us the local
gas pressure $P$, and the original local density can be easily
read out from the slope of axial density profile. Thus the axial
density directly gives us the equation of state ($P$ as a function
of $n$) at the given temperature.

At a given temperature, general thermodynamic relation gives $d
P(r) = n(r) d \mu(r)$ where $\mu(r) \equiv (\mu_\uparrow(r) +
\mu_\downarrow(r))/2$ (Note that $\mu_\uparrow(r) -
\mu_\downarrow(r)$ is independent of $r$). Within the LDA, we can
view the position $r$ as a parameter and thus write $dP(r) = n(r)
d\mu(r)$. Since the the effective chemical potential is
$\mu(r)=\mu_0 - V(r)$, we obtain
\begin{eqnarray} \label{pdV}
dP(r) = -n(r) dV(r)
\end{eqnarray}
This relation can also be obtained by balancing the forces due to
the pressure and trap potential acting on a shell of atoms between
$r$ and $r+dr$. For a harmonic trap,  $V(r)=(\alpha/2)r^2$, we
therefore get $dP (r) = - n(r) \alpha r dr$ and by integration
\begin{eqnarray}
P(r) = \alpha \int_r^{\infty} dr' r' n(r') \label{p-final}
\end{eqnarray}

On the other hand, the axial density at $z$ is defined by
\begin{eqnarray} \label{axi}
n_{\rm axi}(z) = \int_0^{\infty} 2\pi \rho d\rho n(r)
\end{eqnarray}
where  $\rho^2 = x^2 + y^2 = r^2 - z^2$. This can be re-written as
\begin{eqnarray} \label{axi-final}
n_{\rm axi} (z) = 2 \pi \int_z^{\infty} dr r n(r)
\end{eqnarray}
Comparing Eqs.(\ref{p-final})-(\ref{axi-final}) gives
\begin{eqnarray}
P(z) = \frac{\alpha}{2\pi} n_{\rm axi}(z)
\end{eqnarray}
The axial density is thus a direct measure of
 the gas pressure. Also, by
taking the derivative of Eq.(\ref{axi-final}) with respect to $z$,
we get
\begin{eqnarray} \label{daxi}
\frac{d}{dz}n_{\rm axi}(z) &=& -2\pi z n(z)   \ .
\end{eqnarray} Hence
(see also Ref. \cite{pao2})
\begin{eqnarray} \label{den-axi}
n(z) = -\frac{1}{2\pi z}\frac{d}{dz}n_{\rm axi}(z)
\end{eqnarray}
The axial density thus also gives us the original local density
profile \cite{note}.

\begin{figure*}[tbh]
\begin{center}
\includegraphics[width=6.5in]{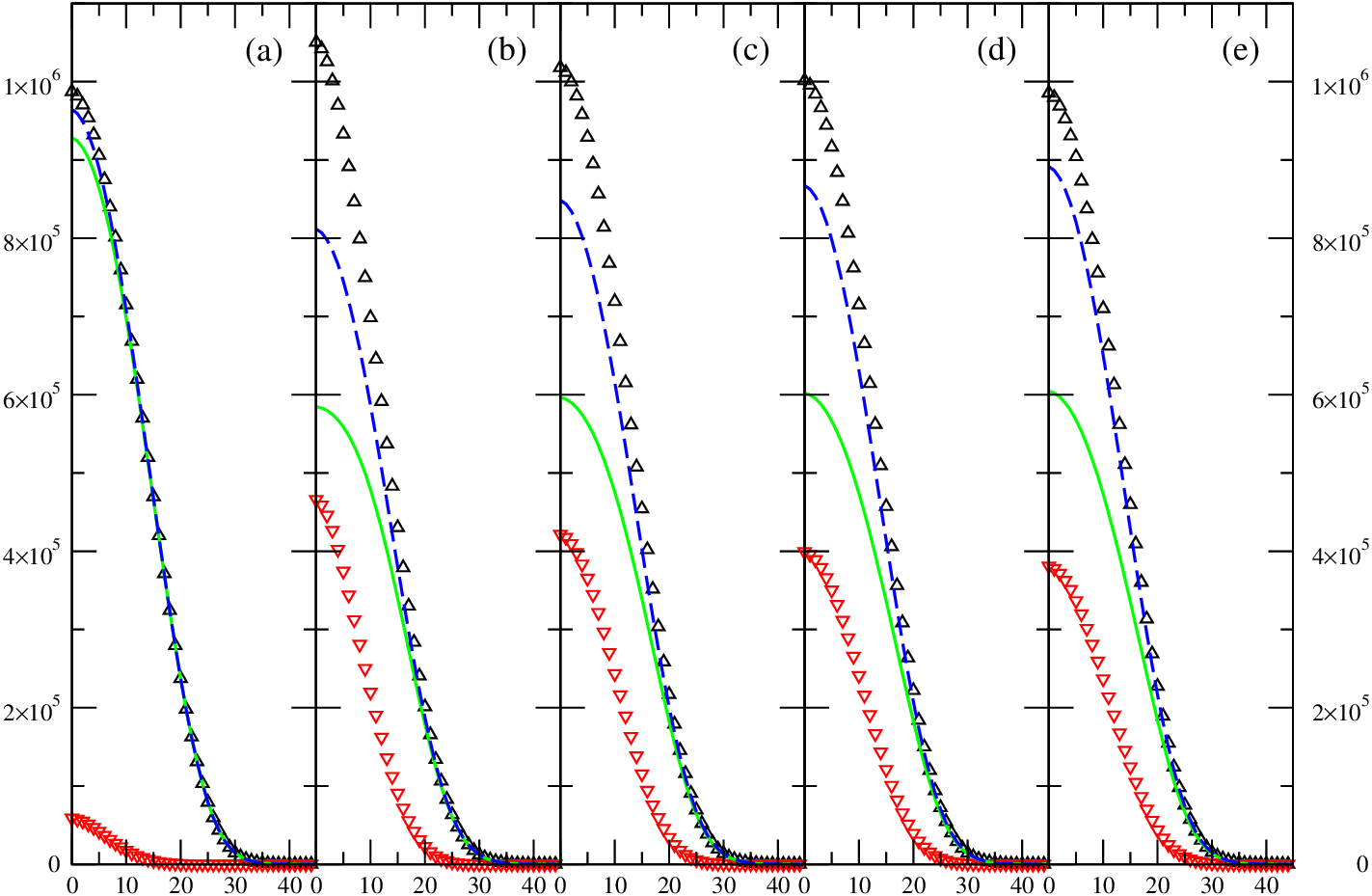}
\end{center}
 \caption{(Color online) Axial density profiles of fermions of spin up and
 down. Labels of horizontal and vertical axes are respectively
 dimensionless $z/\ell$ and $n_{\rm axi}(z)\ell$. Up-spin, down-spin, and excess fermion are represented by
 up-triangle, down-triangle, and solid line, respectively. The dashed line shows the
 trapped non-interacting fermion fit to the wing of spin-up
 fermion profile.
Parameters are the same as in Figs \ref{density.eps} and
\ref{column.eps}.}
 \label{axial.eps}
\end{figure*}

Thus the axial density alone gives us the relation between the gas
pressure and density at a given temperature. For completeness, we
show also the corresponding axial density profiles in Fig.
\ref{axial.eps}.

[For an axially symmetric trap with potential $V(\vec r) =
\frac{1}{2} (\alpha_z z^2 + \alpha \rho^2 )$, the corresponding
relations are: $P(z) =  \frac{\alpha}{2 \pi} n_{\rm axi} (z)$ and
$n(0,0,z) = - \frac{\alpha}{2 \pi \alpha_z} \frac{1}{z}
  \frac { d n_{\rm axi} (z) }{dz}$.]

To conclude, we have studied a two-component interacting Fermi gas
with interaction induced by a Feshbach resonance above the
transition temperature. The density profiles show strong
modifications due to the effect of this interaction.

We acknowledge the support from the Academia Sinica and the
National Science Council of Taiwan under Grant Nos.
NSC94-2112-M-001-002 and  NSC95-2112-M-001-054.

\begin{thebibliography}{}
\bibitem{note}
The relationship between these two densities in Eq.(\ref{den-axi})
can also be shown via the Abel transform and its inverse
\cite{bracewell} within the LDA. The axial density $n_{\rm
axi}(z)$ can also be written as
\begin{displaymath}
n_{\rm axi}(z) \equiv \int_{-\infty}^\infty dy n_{\rm col}(\rho =
\sqrt{y^2+z^2}) = \int_z^\infty d\rho
\frac{2\rho}{\sqrt{\rho^2-z^2}} n_{\rm c}(\rho)
\end{displaymath}
Comparing the inverse Abel transform of this equation with the
definition of the column density in Eq.(\ref{transform-column}),
we obtain Eq.(\ref{daxi}).
\end{thebibliography}
\end{document}